\documentclass[%
 reprint,
nofootinbib,
 amsmath,amssymb,
 aps,
showkeys,
]{revtex4-2}
\usepackage{graphicx}
\usepackage{dcolumn}
\usepackage{bm}
\usepackage{hyperref}

\usepackage{xcolor}
\usepackage{xspace}
\usepackage[capitalize]{cleveref}
\usepackage[normalem]{ulem}

\newcommand{\RE}{RE\xspace}
\newcommand{\REs}{REs\xspace}
\newcommand{\TQ}{TQ\xspace}
\newcommand{\CQ}{CQ\xspace}

\newcommand{\SPARC}{SPARC\xspace}
\newcommand{\REMC}{REMC\xspace}
\newcommand{\DREAM}{DREAM\xspace}
\newcommand{\PRD}{PRD\xspace}
\newcommand{\TRANSP}{TRANSP\xspace}
\newcommand{\DIIID}{DIII-D\xspace}
\newcommand{\NIMROD}{NIMROD\xspace}
\newcommand{\ASCOT}{ASCOT5\xspace}
\newcommand{\GEANT}{GEANT\xspace}
\newcommand{\MHD}{MHD\xspace}
\newcommand{\COMSOL}{COMSOL\xspace}

\newcommand{\VV}{VV\xspace}

\newcommand{\FOW}{FOW\xspace}

\newcommand{\A}{A}
\newcommand{\D}{D}
\newcommand{\T}{T}
\newcommand{\n}{n}
\newcommand{\m}{m}
\newcommand{\Ip}{I_\mathrm{p}}
\newcommand{\Itot}{\Ip}

\newcommand{\Ire}{I_\mathrm{RE}}

\newcommand{\order}[1]{O(\mathrm{#1})}
\newcommand{\Iremc}{I_\mathrm{coil}}

\newcommand{\Bo}{B_0}
\newcommand{\Ro}{R_0}
\newcommand{\Pfus}{P_\mathrm{fus}}
\newcommand{\Wmag}{W_\mathrm{mag}}
\newcommand{\li}{\ell_\mathrm{i}}
\newcommand{\ppar}{p_\parallel}
\newcommand{\vpar}{v_\parallel}
\newcommand{\db}{\delta B / B}
\newcommand{\tTQ}{t_\mathrm{TQ}}

\newcommand{\vteo}{v_\mathrm{te}}
\newcommand{\SI}[2]{#1\,\mathrm{#2}}
\newcommand{\bN}{b_\mathrm{N}}
\newcommand{\psiN}{\psi_\mathrm{N}}
\newcommand{\q}{q}

\newcommand{\mysection}[1]{\noindent\emph{#1.---}}
\newcommand{\replace}[2]{\sout{#1\unskip}\textcolor{red}{#2\unskip}}
\renewcommand{\replace}[2]{#2\unskip}

\newcommand{\add}[1]{\textcolor{red}{#1}}
\renewcommand{\add}[1]{#1\unskip}
\newcommand{\sub}[1]{\sout{#1}}
\renewcommand{\sub}[1]{\unskip}

\newcommand{\nfadd}[1]{\textcolor{red}{#1}}
\renewcommand{\nfadd}[1]{#1\unskip}
\newcommand{\nfsub}[1]{\sout{#1}}
\renewcommand{\nfsub}[1]{\unskip}

\begin{document}

    \title{\add{Modeling the} complete prevention of \add{disruption-generated} runaway electron beam formation with a passive 3D coil \add{in SPARC}}

    \newcommand{\iPSFC}{1}
\newcommand{\iFiatLux}{2}
\newcommand{\iChalmers}{3}
\newcommand{\iIPP}{4}

\author{RA~Tinguely$^\iPSFC$}
\author{VA~Izzo$^\iFiatLux$}
\author{DT~Garnier$^\iPSFC$}
\author{A~Sundstr\"{o}m$^\iChalmers$}
\author{K~S\"{a}rkim\"{a}ki$^\iIPP$}
\author{O~Embr\'{e}us$^\iChalmers$}
\author{T~F\"{u}l\"{o}p$^\iChalmers$}
\author{RS~Granetz$^\iPSFC$}
\author{M~Hoppe$^\iChalmers$}
\author{I~Pusztai$^\iChalmers$}
\author{R~Sweeney$^\iPSFC$}
\affiliation{$^\iPSFC$Plasma Science and Fusion Center, Massachusetts Institute of Technology, Cambridge, MA 01239, USA}
\affiliation{$^\iFiatLux$Fiat Lux, San Diego, CA 92101, USA}
\affiliation{$^\iChalmers$Department of Physics, Chalmers University of Technology, SE-41296 G\"{o}teborg, Sweden}
\affiliation{$^\iIPP$Max Planck Institute for Plasmaphysics, 85748 Garching, Germany}


    \begin{abstract}
    The potential formation of multi-mega-ampere beams of relativistic ``runaway'' electrons (\REs) during sudden terminations of tokamak plasmas poses a significant challenge to the tokamak's development as a fusion energy source. Here, we use state-of-the-art modeling of disruption magnetohydrodynamics coupled with a self-consistent evolution of \RE generation and transport to show that a non-axisymmetric in-vessel coil will \emph{passively} prevent \RE beam formation \add{during disruptions} in the \SPARC tokamak, a compact, high-field, high-current device capable of achieving a fusion gain $Q>2$ in deuterium-tritium plasmas.
\end{abstract}


    \maketitle 
    \mysection{Introduction}\label{sec:intro}%
    The tokamak, a magnetic confinement fusion device, is perhaps the most promising concept extrapolating to a future fusion power plant. Yet it is not without an Achilles' heel: A disruption is the sudden, often unplanned termination of a plasma discharge caused by a host of instabilities \cite{Hender2007,Sweeney2020}. While instability limits can be avoided so that disruptions occur with low probability, tokamaks must be designed to mitigate disruptions' deleterious effects.
    
    This paper focuses on the mitigation of so-called ``runaway'' electrons (\REs). The sudden cooling of a plasma during the disruption's thermal quench (\TQ) leads to an increase in the plasma resistivity $\propto \T^{-3/2}$, with $\T$ the plasma temperature. In an attempt to maintain the pre-disruption plasma current \replace{$\Ip$}{}, 
    a large electric field is induced, accelerating electrons continuously against a decreasing collisional friction $\propto v^{-2}$, with $v$ the electron speed. The \RE population is exponentially amplified by the same avalanche process proposed for atmospheric breakdown during lightning \cite{gurevich1992runaway}, and the resulting \RE beam, carrying currents \nfsub{$\order{MA}$} \nfadd{of order MA} and with individual electron energies \nfsub{$\order{MeV}$} \nfadd{of order MeV}, can be destructive if not avoided or mitigated \cite{Lehnen2015,Matthews2016}.
    
    The most studied mitigation method is massive material injection (gas or shattered pellet) either before or after \RE beam formation \cite{Reux2015,Shiraki2018,PazSoldan2019,Reux2021}. \RE avoidance has also been pursued via applied 3D magnetic fields, with the resulting stochasticity transporting \REs out of the plasma \nfadd{\cite{Catto1991,Lehnen2008,Commaux2011,Chen2016,Gobbin2018,Chen2018,Lin2019,Munaretto2020}}. While initial results are promising, fully stochastic fields have not yet been achieved \cite{Yoshino2000,Smith2013}, and this is still an \emph{active} mitigation scheme.
    
    In this work, we demonstrate complete prevention of \add{disruption-generated} \REs in the \SPARC tokamak \cite{Creely2020} via a \emph{passive} mitigation strategy. A conducting coil with 3D structure -- the \RE Mitigation Coil (\REMC) -- is energized by the disruption-induced voltage, and the resulting magnetic field stochasticity causes electrons to be lost faster than a \RE beam can form \cite{Boozer2011,Smith2013}. In this Letter, we focus on the worst-case scenario, \replace{i.e.}{} a disruption of the ``primary reference discharge'' \nfadd{(\PRD)}: a deuterium-tritium, H-mode plasma with \replace{major and minor radii $\Ro = \SI{1.85}{m}$ and $a = \SI{0.57}{m}$}{major/minor radii $\Ro = \SI{1.85}{m}$/$a = \SI{0.57}{m}$}, toroidal magnetic field $\Bo = \SI{12.2}{T}$, plasma current $\Ip = \SI{8.7}{MA}$, volume-averaged density $\langle\n\rangle \approx \SI{3\times10^{20}}{m^{-3}}$ and temperature $\langle\T\rangle \approx \SI{7}{keV}$, and fusion power $\Pfus = \SI{140}{MW}$. \replace{Plasma profiles used in this work come from the self-consistent \TRANSP simulations in \cite{Rodriguez-Fernandez2020}.}{} 
    We assess the efficiency of \REMC by first computing its vacuum magnetic field perturbation, which is then included in 3D magnetohydrodynamics (MHD) modeling; in the resulting fields, \RE orbits are traced to calculate transport coefficients for a fluid-kinetic solver of \RE generation and transport.
    
\mysection{Vacuum field modeling \replace{with COMSOL}{}}\label{sec:comsol}%
    All toroidally continuous, conducting structures in the \SPARC tokamak are modeled with the 3D finite element code \COMSOL~\cite{comsol}: the central solenoid, double-walled vacuum vessel (\VV), passive stability plates, and poloidal field, divertor, and vertical stability (VS) coils \cite{Creely2020,Sweeney2020}. The \nfadd{copper-based} $\n=1$ \REMC is situated at the outboard \VV wall with two vertical legs avoiding ports and two horizontal legs near the VS coils \replace{inside the \VV}{} \add{(see Fig. 12(a) of \cite{Sweeney2020})}. An elliptical toroid of uniform current density approximates the post-\TQ redistribution of plasma current, and a linear decay over $\SI{3.2}{ms}$ simulates the fastest expected current quench (\CQ) from empirical scalings \cite{Sugihara2004,Sweeney2020}. The mutual inductance between the plasma and \REMC leads to a nearly linear ramp in \REMC current\nfsub{ with}\nfadd{; the} peak value $\Iremc = \SI{590}{kA}$ \nfadd{is the maximum achievable current (driven passive-inductively) using this particular coil placement and resistivity}. The result is \replace{time- and}{} $\Ip$-dependent vacuum magnetic perturbations, along with the magnetic energy available for dissipation during the \CQ.
    
    \begin{figure}[h!]
        \centering
        \includegraphics[width=\columnwidth]{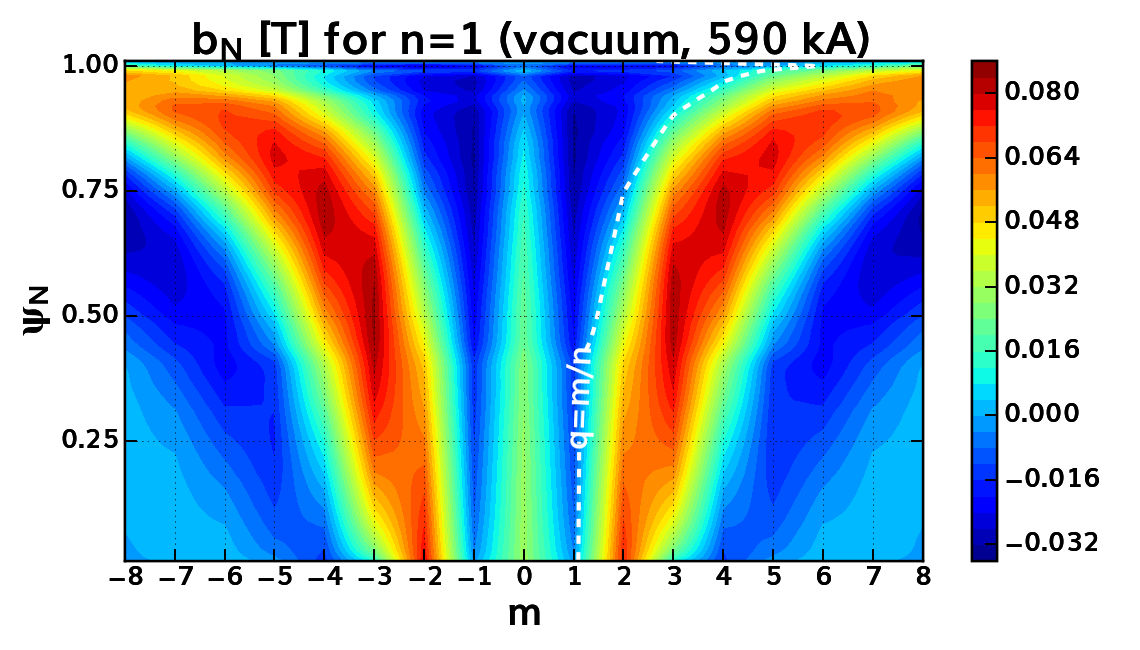}
        \caption{\nfadd{Poloidal spectrum of $\n=1$ vacuum magnetic perturbations from the \REMC at maximum current: the amplitude $\bN$~(T) normal to (normalized) flux surface $\psiN$ vs poloidal mode number $\m$, with safety factor profile $\q$ overlaid (dashed).}}
        \label{fig:polspec}
    \end{figure}
    
    \nfadd{
    \Cref{fig:polspec} shows the poloidal ($\m$) spectrum in straight field-line coordinates of the dominant $\n=1$ vacuum magnetic field - specifically the component normal to the \PRD's flux surfaces - at maximum $\Iremc$. As expected, the spectrum is symmetric around $\m=0$, and the position of peak amplitude moves inward (farther from the \REMC) with decreasing $\vert\m\vert$. While the vacuum field does not appear to be resonant with this particular safety factor profile $\q$, field distributions like this are expected to drive the strongest kink response of the plasma and thereby the strongest resonant fields \cite{PazSoldan2014}. This is evident in the results of \MHD modeling in the next section.
    }
    
    
    \begin{figure*}[t!]
    \centering
    \includegraphics[width=1.0\textwidth]{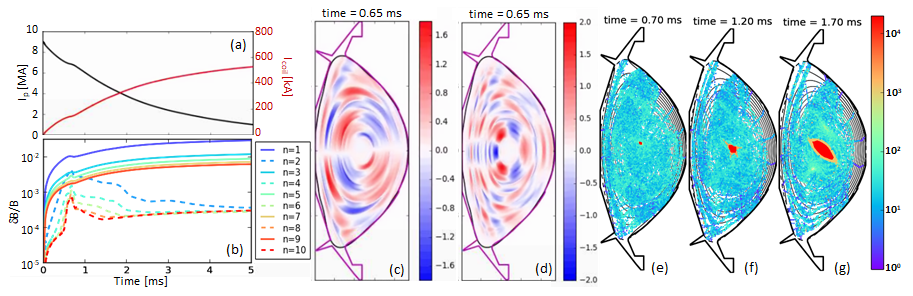}
    \caption{\NIMROD simulation results for the $\n=1$ \REMC: (a)~plasma current and coil current vs time; (b)~amplitude of $\n=1{-}10$ modes in units of $\db = \sqrt{\Wmag(\n)/\Wmag(\n=0)}$; (c)~$\n=1$ and (d)~$n=2$ toroidal current densities $\mathrm{(MA/m^2)}$ at the time of nonlinear mode saturation (with black line the simulation boundary and magenta line the \SPARC first wall shape); \nfadd{(e--g)~magnetic field line Poincar\'{e} plots on top of poloidal flux contours, where colors represent the number of toroidal transits before exiting or reaching the maximum integration length of 20,000}. The empty regions near the boundary in (e--g) are artifacts of short field lines that quickly exit the volume.   
    }
    \label{fig:NIMROD}
    \end{figure*}

\mysection{3D MHD modeling \replace{with \NIMROD}{}}\label{sec:nimrod}%
    The \NIMROD 3D MHD code \cite{sovinec:2004} is used to model only the \CQ phase of the \replace{\SPARC}{} disruption, with external 3D magnetic fields from \COMSOL \replace{ramped in proportion to the plasma current decay}{}. Only the surface normal components of the $\n=1{-}10$ toroidal Fourier modes are applied at the location of the \NIMROD simulation boundary. \NIMROD has previously been run with applied time-changing boundary fields to model the application of resonant magnetic perturbations in the \DIIID tokamak \cite{Izzo2008}. The present model differs only in that the amplitudes' time-dependence is not specified \emph{a priori}; rather, it is calculated in proportion to the \replace{plasma current}{$\Ip$} decay as the simulation evolves, using the value of the maximum \REMC current ($\SI{590}{kA}$) from \COMSOL: 
    \begin{equation*}
        \Iremc(t)=\max(\Iremc) \times \left[1-\Ip(t)/\Ip(t=0)  \right].
    \end{equation*}

    
    Beginning with an $\SI{8.7}{MA}$ steady-state \SPARC equilibrium \cite{Rodriguez-Fernandez2020}, a \CQ is initiated \replace{by means of}{via} an artificially induced \TQ wherein the coefficient of perpendicular thermal conductivity is set to a large value, $\SI{4\times10^4}{m^2/s}$, for only the first $\SI{0.045}{ms}$ of the simulation, until essentially all of the thermal energy is lost; then it is reduced to a more modest \replace{value of}{} $\SI{2}{m^2/s}$. A more realistic \TQ induced by impurity radiation has \replace{been shown to produce}{shown} significant losses of \REs in SPARC~\cite{Sweeney2020}, but the \nfadd{present artificial-TQ method} conservatively bypasses any \TQ MHD and only considers the \CQ \replace{phase}{}.%
    \nfadd{\footnote{
        \nfadd{Though not shown here, a \NIMROD simulation including the artificial \TQ and \emph{not} including the \REMC results in \emph{no} \CQ MHD, meaning that the MHD observed here is solely due to the coil.}
    }} 
    Once the plasma is cold and the thermal diffusivity is reduced, a balance of Ohmic heating and radiation from $4.8\times10^{21}$ atoms of added Neon determines the \CQ temperature, \replace{which falls between $\T = \SI{3}{eV}$ at the plasma edge and $\SI{8}{eV}$ in the core}{$\T \approx \SI{3{-}8}{eV}$ from the plasma edge to core}. 
    
    As \replace{the plasma current}{$\Ip$} begins to decay and the external fields grow, odd mode numbers are predominantly perturbed by the \REMC (\cref{fig:NIMROD}(a,b)). \replace{But}{} Nonlinear growth of every mode is quickly excited, saturating around $t \approx \SI{0.65}{ms}$, with $\n=2$ being the second largest mode after $\n=1$ \replace{at the time of saturation}{}. The $\n=1$ and $2$ saturated, nonlinear modes include many poloidal \nfsub{($\m$)} harmonics of comparable amplitude, with \nfsub{the $\n=2$ mode including a prominent $\m=3$ component} \nfadd{$\m/\n = 1/1$, $2/1$, and $3/2$ modes prominent} near the core (\cref{fig:NIMROD}(c,d)). The growth of the these modes is responsible for a redistribution of the current density profile, producing a small inflection in the \replace{total plasma current}{$\Ip$} decay as the plasma internal inductance $\li$ decreases. Near-complete destruction of magnetic flux surfaces occurs shortly after \replace{the time of nonlinear mode saturation}{} ($t\approx\SI{0.7}{ms}$), but a small island\nfadd{, with width of order cm,} \nfsub{$\order{cm}$} reappears in the core within $t=\SI{0.5}{ms}$ and continues to slowly grow \replace{in the core}{} even as the amplitude of the external fields further increases (\cref{fig:NIMROD}(e--g)). Therefore, global magnetic field stochasticity is achieved temporarily by a combination of the applied external fields and the excitement of nonlinear MHD mode growth.
    

        \begin{figure*}[t!]
        \centering
        \includegraphics[width=1.0\textwidth]{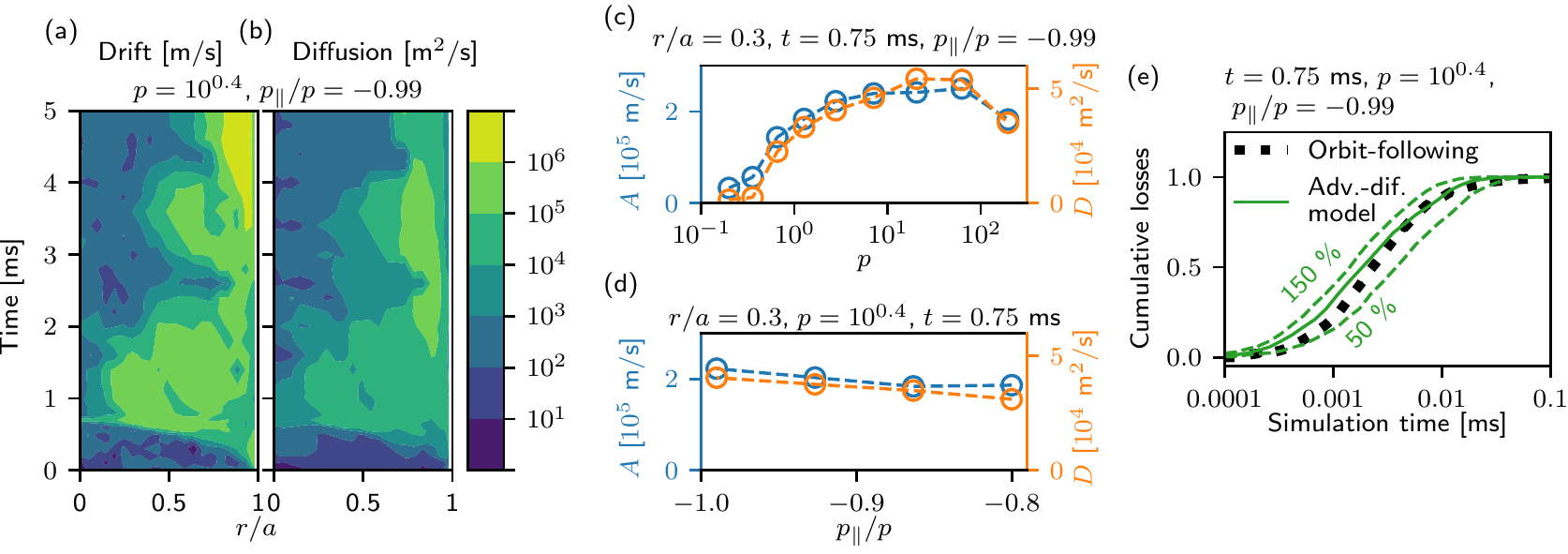}
        \caption{(a--d):~Transport coefficients as a function of (a,b)~time and radius, (c)~normalized momentum, and (d)~pitch.
        The circles in (c) and (d) indicate the coordinates at which the coefficients were evaluated.
        (e)~The cumulative fraction of markers lost as a function of simulation time computed with the advection--diffusion model (green) compared to the equivalent orbit-following \nfadd{\ASCOT} simulation (dotted black).
        Dashed green lines in (e) correspond to cases when the coefficients are scaled by factors 150\% / 50\% of the nominal value (solid green). Fixed values are given above each plot.
        }
        \label{fig:transport}
    \end{figure*}

\mysection{Evaluation of transport coefficients \replace{with \ASCOT}{}}\label{sec:ascot}%
    The orbit-following code \ASCOT~\cite{hirvijoki2014ascot} is used to compute the advection \replace{(drift)}{} and diffusion coefficients \replace{}{($\A,\D$)} characterizing the radial transport of electrons in the 3D stochastic fields from \NIMROD. Markers are traced in stationary times-slices, initialized with different radial positions~$r$, momenta~$p$ (normalized by the electron mass~$m_e$ and speed of light~$c$), and pitches~$\ppar/p$ \replace{w.r.t. the magnetic field}{}. \replace{The evaluated coefficients are stored as 4D $(r/a,p,\ppar/p,t)$ tables.}{} The transport is assumed to be dominated by the field stochasticity, and therefore collisions and electric field acceleration are not included \footnote{The impact of magnetic islands on transport are not considered in \ASCOT calculations, but are incorporated in some \DREAM simulations.}. 
    This numerical scheme is used because the alternative, Rechester-Rosenbluth diffusion~\cite{Rechester-Rosenbluth_PRL1978}, does not accurately portray transport in fields perturbed by external coils~\cite{papp_2015,sarkimaki2016advection} and also omits finite-orbit-width (\FOW) effects that can reduce \RE transport ($\propto p^{-1}$) at sufficiently high energy \cite{Hauff_2009,Myra_1992}.
    
    The method of evaluating the transport coefficients is adopted from an earlier scheme \replace{by Boozer and Kuo-Petravic}{} \cite{boozerpetravic1981} and \replace{is}{} described in \cite{sarkimaki2016advection}.
    At the beginning of the orbit-following simulation, markers are assigned $(r/a,p,\ppar/p)$ values and random toroidal positions.
    They are then traced for $\SI{0.02}{ms}$ or until they pass the separatrix, and each time a marker crosses the outer midplane its position is recorded.
    For markers that remain confined, the time-evolution of the recorded radial positions is used to estimate \replace{the advection $\A$ and diffusion $\D$ coefficients}{$\A$ and $\D$}.
    For markers that are lost, the \replace{distribution of the loss-time is}{loss-times are} used to fit the \replace{so-called}{} ``first passage time distribution,'' with $\A$ and $\D$ fitted parameters.
    Finally, the mean value is taken from the marker-specific coefficients to represent $\A$ and $\D$ at that position in phase space \replace{$(r/a,p,\ppar/p,t)$}{}.
    
    Samples of evaluated coefficients are shown in \cref{fig:transport}(a--d).
    \replace{What is observed holds generally for other fixed values unless otherwise stated.}{}
    The time-interval over which stochastic fields reach and persist in the core ($t\approx\SI{0.7{-}2.0}{ms}$) is reflected in the transport coefficients. 
    During this time, \replace{the advection is}{} \nfsub{$\A\sim\order{\SI{10^5}{m/s}}$} \nfadd{$\A\sim \SI{10^5}{m/s}$} and \replace{the diffusion}{} \nfsub{$\D\sim\order{\SI{10^4}{m^2/s}}$} \nfadd{$\D\sim\SI{10^4}{m^2/s}$}. 
    The transport grows initially with momentum \replace{, as $\vpar \to c$}{}, but decreases for $p>10^{2}$ \nfadd{($E>\SI{50}{MeV}$)}, likely due to \FOW effects.
    Likewise, the pitch dependency is linear except when the electron energies are high and \FOW effects again become relevant. \nfadd{However, as described in the next section, most \REs are expected to attain energies ${<}\SI{20}{MeV}$, thereby minimizing these effects.}
    
    \Cref{fig:transport}(e) shows \replace{the accumulation of}{accumulated} losses when the advection-diffusion equation is solved numerically using the evaluated coefficients. Here, the initial population consists of electrons \replace{that are}{} uniformly distributed radially.
    To confirm the validity of modeling \RE transport as an advection-diffusion process, we \replace{also}{} show the result of an equivalent \ASCOT simulation where the same population \replace{was}{is} traced in the 3D field until the losses saturated.
    Good agreement is seen between the model and simulation; as expected, scaling the coefficients by 50\% (150\%) underestimates (overestimates) the transport.

        \begin{figure*}[t!]
        \centering
        \includegraphics[width=1.0\textwidth]{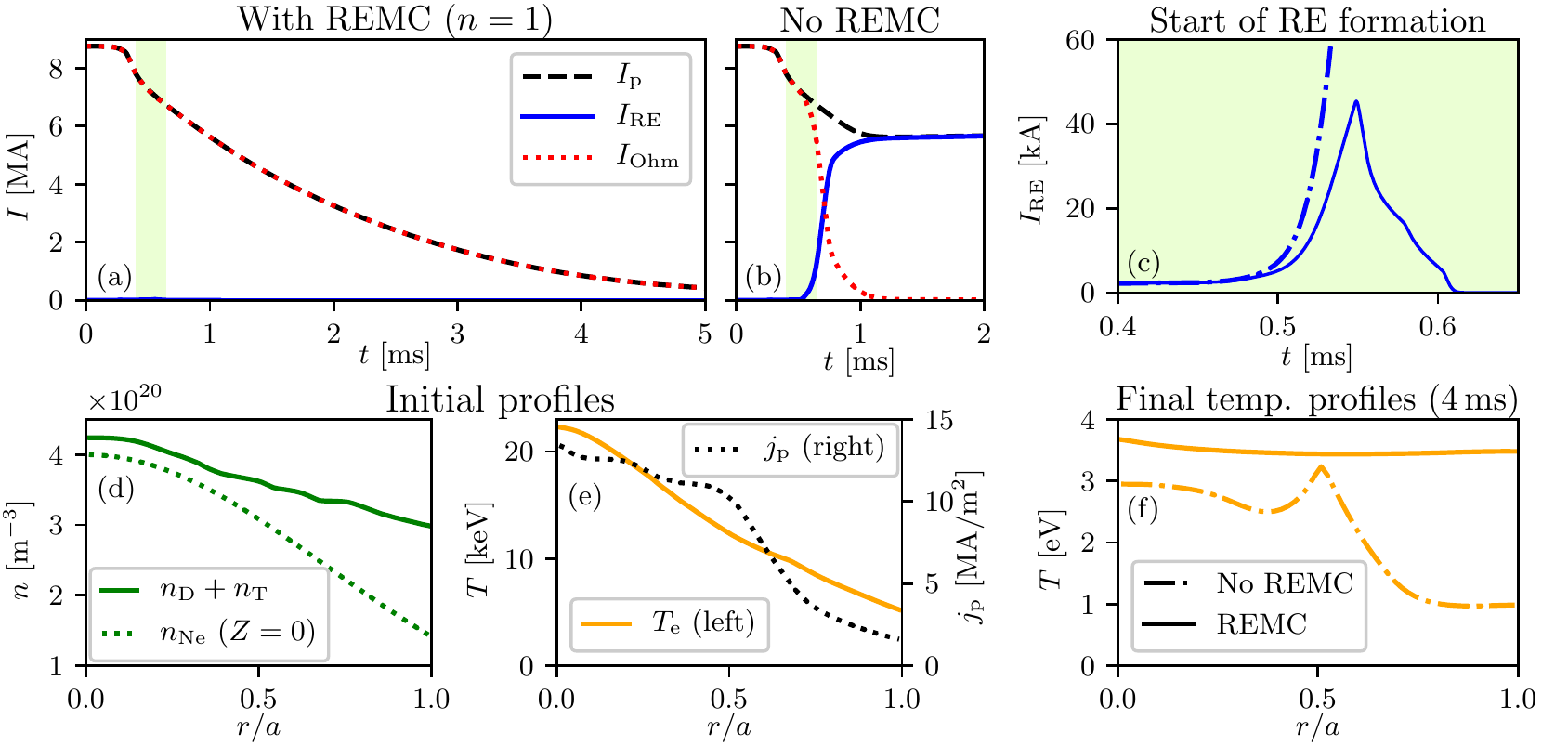}
        \caption{\DREAM simulations of the current evolution with the \REMC (a)~active and (b)~inactive, showing the total plasma current (dashed) with Ohmic (dotted) and \RE (solid) components. 
        (c)~An expanded view of the \RE current in the shaded intervals of (a,b) with the \REMC active (solid) and inactive (dot-dashed).
        Initial profiles of the (d)~deuterium + tritium and neutral neon densities and (e)~electron temperature and total current density. 
        (f)~Final temperature profiles at $t=\SI{4}{ms}$.
        }
        \label{fig:DREAM}
    \end{figure*}

\mysection{Runaway electron evolution \replace{with \DREAM}{}}\label{sec:dream}%
    The 1D radial-transport solver in the \RE modeling framework \DREAM~\cite{Hoppe_DREAM-2021,Svenningsson2021} is used to consistently evolve the electric field and \RE generation -- including Dreicer, hot-tail, avalanche, tritium beta decay, and inverse Compton scattering source terms -- throughout the disruption.
    \DREAM's fluid \RE transport model, based on~\cite{Svensson_JPP2021}, simulates the effects of the \REMC: time-dependent transport coefficients from \ASCOT are mapped to \replace{the total current}{} $\Itot(t)$ -- as in \NIMROD (\cref{fig:NIMROD}(a)) -- and interpolated in the \DREAM simulation based on the instantaneous value. A comparison of the total \RE currents in \cref{fig:DREAM}(a) and (b), with and without the \REMC, shows that the \REMC effectively inhibits \RE formation.
    As the \CQ begins at $t\approx\SI{0.5}{ms}$, a ${\sim}\SI{2}{kA}$ hot-tail seed rapidly grows (\cref{fig:DREAM}(c)). With the \REMC \emph{inactive}, over $60\%$ ohmic-to-\RE current conversion is observed, \add{similar to that seen in previous simulations \cite{Sweeney2020};}
    \sub{and} average/maximum \RE energies attained are ${\sim}\SI{8}{MeV}/\SI{69}{MeV}$, \add{with 90\% of \REs having energies ${<}\SI{17}{MeV}$}.%
    \nfadd{\footnote{\nfadd{Synchrotron and bremsstrahlung power loss mechanisms are included in \DREAM. Drift orbit losses are not included in \DREAM explicitly, but rather implicitly through the transport coefficients calculated with \ASCOT.}}}
    However, with the \REMC \emph{active}, the \RE dissipation rate due to transport surpasses the growth rate at $t\approx\SI{0.55}{ms}$, and the \RE current rapidly diminishes after a short transient $\Ire\sim\SI{50}{kA}$.
    
    
    These \DREAM simulations, performed in the fully fluid mode, are initiated with the plasma profiles shown in \cref{fig:DREAM}(d) and (e). In the first $\tTQ=\SI{0.27}{ms}$ of the simulation, an exponential temperature decrease is prescribed for the \TQ, corresponding to an initial drop in the central temperature from ${\sim}\SI{20}{keV}$ to $\SI{100}{eV}$. Afterward, the temperature evolution is calculated from power balance, including Ohmic heating, radiation losses, and heat diffusion due to magnetic field perturbations
    , assuming $\vert\db\vert=3\times10^{-3}$ (\cref{fig:NIMROD}(b)). 
    The effect of the \REMC on the final temperature profile \replace{(at $t=\SI{4}{ms}$)}{} is shown in \cref{fig:DREAM}(f); these agree with \NIMROD results within a factor of two. 
    
    The maximum \RE current is highly sensitive to the magnetic energy available for dissipation, determined by the \replace{toroidally closed}{} conducting geometry around the plasma. The magnetic boundary surface in \DREAM is placed at $r/a=1.19$, enclosing a poloidal magnetic energy of $\SI{52.8}{MJ}$, slightly larger than the $\SI{52.3}{MJ}$ calculated in \COMSOL. Note that small changes in this energy can lead to significant changes in the peak current, e.g.\ $\Delta r/a \approx +1\% \to \Delta\Ire \approx +50\%$. The \DREAM simulations also use a finite resistive-wall time of $\SI{50}{ms}$, consistent with \SPARC's double-walled \VV, allowing \replace{some}{} additional magnetic energy to diffuse into the domain. 
    
    
    In \SPARC, the \TQ duration is estimated to vary between $\tTQ=\SI{0.1{-}1}{ms}$ \cite{ITERPhysicsBasis1999,Sweeney2020}, with hot-tail generation increasing with decreasing $\tTQ$. For $\tTQ=\SI{0.1}{ms}$, a hot-tail \RE current forms, reaching ${\sim}\SI{1.2}{MA}$ at $t\approx\SI{0.5}{ms}$, but then dissipates quickly by $t\approx\SI{0.55}{ms}$. In this simulation, \RE transport during the \TQ is estimated using a (conservative) \RE diffusivity $(c/\vteo)a^2/\tTQ\approx \SI{9000}{m^2/s}$, with $\vteo$ the electron thermal speed at $\SI{20}{keV}$.       
    
    The robustness of the result is further tested by artificially reducing transport coefficients in our baseline case by $90\%$, yielding a \RE current spike of only $\SI{150}{kA}$ which dissipates within $\SI{0.1}{ms}$ of the peak.
    In another series of \NIMROD+ \ASCOT+ \DREAM simulations, the induced \REMC current is halved, yielding a transient \replace{\RE peak of}{} $\Ire\approx\SI{2}{MA}$ at $t=\SI{0.75}{ms}$, but the \RE beam is dissipated completely by $t=\SI{0.87}{ms}$. The striking difference between the maximum \replace{\RE currents}{$\Ire$} in the two cases -- i.e. $(\A,\D)\times 10\%$ vs $\Iremc \times 50\%$ -- is due to a delayed onset and shallower penetration of magnetic perturbations in the latter. 
    
    Lastly, we address the potential effect of a \replace{}{core} magnetic island \replace{in the center of the plasma}{} (\cref{fig:NIMROD}(e--g)), where \REs could be confined. Simulations with $\A=0$ and $\D=0$ inside $r\leq\SI{4}{cm}$ yields $\Ire \approx \SI{2}{MA}$ within that region. However, removing all \RE transport is excessively conservative; not only would drifts due to the induced electric field~\cite{Guan2010} cause some \REs to escape the island, but the high current density -- corresponding to safety factors $q \ll 1$ and \replace{an internal inductance}{} $\li \gg 1.5$ -- will very likely never develop due to the self-regulating kink instability~\cite{Cai_2015,PazSoldan2019};
    the self-consistent modeling of which is outside the scope of this study.   
    
\mysection{Discussion}\label{sec:discussion}%
    These promising results have motivated our team to move forward with the engineering design of the $\n=1$ \REMC. Here, we discuss practical requirements of the coil and its impact on tokamak operation. 
    
    The \REMC will experience various electromechanical stresses and forces. The $\n=1$ configuration introduces a large sideways force \replace{}{(${\leq}\SI{15}{MN}$)} due to oppositely directed currents in the two vertical legs\replace{, each of length $L \approx \SI{1.2}{m}$: $F \approx 2\Iremc L \Bo\Ro/(\Ro+a) \leq \SI{15}{MN}$}{}. The present plan is to mount the \REMC close to the vertical stability coils, and structural analysis is underway. Net-force-free $\n=1$ alternatives are also under evaluation. \nfadd{Furthermore, full \RE suppression might be achieved at a lower maximum \REMC current than that specified here; this will be explored both through simulations and as part of the \REMC's experimental program. One option to decrease $\Iremc$ is simply to add an external adjustable resistor to the circuit.}
    
    The \REMC switch will be located outside the \VV for ease of access and radiation protection, but its type has not been decided. The preferred option is a \sub{solid state switch, spark gap, or varistor} \add{Shockley diode} that is passively activated by high voltages unique to the disruption. The absence of any switch (i.e. a continuous coil) or a mechanical switch \replace{that is}{} closed after the \replace{plasma current}{$\Ip$} ramp-up has also been considered, but providing protection during ramp-down without inadvertently causing a disruption appears challenging. Preliminary assessments suggest that the resistance required to keep \replace{the \REMC currents}{$\Iremc$} at acceptable levels during ramp-down would degrade its performance during disruptions. Furthermore, while the induced voltage from common internal $\m/\n=1/1$ kinks (``sawtooth'' crashes) is expected to be negligible,  the impact of H-mode's edge-localized modes on a closed-circuit coil is unknown and currently under investigation.
    
    The mitigation of \REs generated during the low density plasma current ramp-up is a natural extension of the \REMC's function \replace{beyond disruptions}{}. Unfortunately, the high voltage that would passively activate the \REMC during disruptions is not present during a start-up \RE event. Powering the \REMC using a capacitor bank could be considered if the simpler approach of controlling and ramping-down the \RE beam is not possible.
    
    While only the $\n = 1$ \REMC configuration is reported in this Letter, $\n=2$ and $3$ designs have also been considered \cite{Sweeney2020}. However, both \NIMROD and \DREAM simulations indicate reduced and \emph{no} effects of the $\n = 2$ and $3$ coils, respectively, on \RE formation as the induced transport does not extend far enough into the plasma center (\cref{fig:transport}(a,b)), where most \REs are generated. Because the engineering requirements of the $\n=1$ coils are feasible, the $\n=2$ and $3$ designs will not be pursued further.

    \mysection{Summary}\label{sec:summary}%
    In this Letter, a non-axisymmetric in-vessel coil, with a toroidal $\n=1$ geometry, was concluded to be capable of passively preventing post-disruption \RE beam formation in the \SPARC tokamak. With \COMSOL, we modeled the energization of the coil during a realistic \CQ.
    The resulting 3D vacuum magnetic fields were provided as inputs to \NIMROD, which simulated the full \CQ MHD. Using \ASCOT, we evaluated the transport coefficients of fast electrons in \NIMROD's highly stochastic fields, and the self-consistent evolution of the \RE beam, with and without the coil, was simulated with \DREAM.
    
    Near-term work will focus on modeling \RE impacts on and heating of plasma-facing components \replace{with \ASCOT and \GEANT \cite{GEANT4}}{}. A helical \REMC is also under design for installation in the similarly sized \DIIID tokamak \cite{Dunn2020,Weisberg2021}. Experimental validation of this Letter's projected \REMC efficiency and exploration of potential (and unknown) side effects will be hugely influential in the decision to close the switch in \SPARC. 

    \begin{acknowledgments}
    The authors thank M~Greenwald and P~Rodriguez Fernandez for fruitful discussions. This work was supported by Commonwealth Fusion Systems and the Swedish Research Council (Dnr.~2018-03911). We acknowledge the CINECA award, under the ISCRA initiative, for the availability of high performance computing resources and support. This research used resources of the National Energy Research Scientific Computing Center (NERSC), a U.S. Department of Energy Office of Science User Facility operated under Contract No.~DE-AC02-05CH11231. Part of the data analysis was performed using the OMFIT integrated modeling framework \cite{Meneghini_2015}.
\end{acknowledgments} 
    
    
%

\end{document}